\documentclass[12pt]{article}

\usepackage{hyperref}
\usepackage[euler]{textgreek}
\usepackage{scicite}
\usepackage{graphicx}
\usepackage{times}
\usepackage{xcolor}

\newenvironment{sciabstract}{%
\begin{quote} \bf}
{\end{quote}}

\title{Light-based Volumetric Additive Manufacturing in Scattering Resins}

\author
{Jorge Madrid-Wolff$^{1, \dagger}$, Antoine Boniface$^{1, \dagger}$, Damien Loterie$^{2}$, 

\\

Paul Delrot$^{2}$, Christophe Moser$^{1, \ast}$ \\
\\
\normalsize{$^{1}$ Laboratory of Applied Photonics Devices, School of Engineering,} 
\\
\normalsize{Ecole Polytechnique Fédérale de Lausanne, CH-1015, Lausanne, Switzerland
}\\
\normalsize{$^{2}$ Readily3D SA
EPFL Innovation Park, Building A
CH-1015 Lausanne, Switzerland}\\
\normalsize{$^{\dagger}$ These authors contributed equally to this work}\\
\\
\normalsize{$^\ast$To whom correspondence should be addressed; E-mail:  christophe.moser@epfl.ch}
}

\date{}

\begin{document}


\baselineskip24pt


\maketitle


\begin{sciabstract}
3D printing has revolutionized the manufacturing of volumetric components and structures for use in various fields. Owing to the advent of photo-curable resins, several fully volumetric light-based techniques have been recently developed to push further the resolution and speed limitations of 3D printing. However, these new approaches only work with homogeneous and relatively transparent resins so that the incident light patterns used for photo-polymerization are not impacted along their propagation through the material. Herein, we describe a strategy to print in scattering materials. It consists of characterizing how light is distorted by the curable resin and then applying a digital correction to the light patterns to counteract the effect of scattering. Using a tomographic volumetric printer, we experimentally demonstrate the importance of taking light scattering into account when computing the projected patterns and show that our applied correction significantly improves printability, even when the object size exceeds the scattering mean free path of light. 
\end{sciabstract}


\section{Introduction}
Until only a few years ago, the conventional approach in light-based additive manufacturing, or 3D printing, relied on constructing objects by piling one-dimensional voxels  or two-dimensional layers on top of each other. Each layer being formed by the solidification of a photo-curable resist under light irradiation, for example, by either scanning a laser beam point-by-point, namely stereolithography (SLA) \cite{hull1984SLA_patent} or two-photon fabrication \cite{lee2006twoPhotonPolym, farsari2009TwoPhotonFabrication}, or by projecting two-dimensional light patterns, namely digital light processing (DLP) technology. A 100-fold increase in speed in DLP printing was achieved with the CLIP \cite{tumbleston2015continuous} and HARP methods \cite{walker2019rapid}. However, these are still based on a layer-by-layer strategy.
Recently, new methods for additive manufacturing of cm-scale 3D objects in photo-sensitive materials have been proposed, including the holographic display of light patterns \cite{shusteff2017holographic3D}, two-photon polymerization \cite{chu2018centimeterScale2PP}, xolography \cite{regehly2020xolography}, and tomography  \cite{kelly2019volumetric,loterie2020high}. Tomographic additive manufacturing consists of illuminating a volume of transparent and photo-responsive material with a set of light patterns from multiple angles. The cumulative light exposure results in a volumetric energy dose that is sufficient to solidify the material in the desired geometry. The main advantages of tomographic volumetric printing compared to existing methods are its short manufacturing time (down to a few tens of seconds), and its ability to print complex hollow structures without the need for support structures as required in layer-by-layer fabrication systems. However, to achieve a correct three-dimensional light dose deposition in the build volume, the light patterns projected from multiple angles must illuminate the entire build volume. This restricts the printing technique to transparent materials with very little light scattering and/or absorbance. 

The tomographic-based methods are called volumetric because they depart from the sequential fabrication process to construct the object in a true three-dimensional fashion. 
This technique enables printing objects whose scale is of the order of 2 cm x 2 cm x 2 cm with a typical resolution of up to 80 micrometers \cite{kelly2019volumetric,loterie2020high}. As any projection-based printing system, the final resolution of the printed structure is, in theory, determined by the effective pixel size of the DLP projector at the center of the build volume. Other works have expanded the applicability of tomographic volumteric printing by reducing optical aberrations inherent to the experimental setup \cite{orth2021corrRayDistTomo3DPrint}, polymerizing thiol-ene photoresins \cite{cook2020thiol-ene}, and compensating for non-simultaneous polymerisation \cite{loterie2020high, chung2020schlieren}.
Most problematically, light may also have unwanted interactions with the material itself. These are frequently overlooked as most resins used up to now are transparent and almost non-absorptive, but must be considered when operating with more complex materials. This is the case when printing in scattering material like biological inks, hydrogels, or composite resins. These materials are of utmost interest for many applications including bioprinting \cite{murphy20143d}, medical devices \cite{liaw2017current}, customized implants \cite{dawood20153d} and even jewelry \cite{yap2014additive}. Scattering of light represents a major problem for any light-based 3D printer as it severely limits their resolution \cite{you2020mitigating, zakeri2020comprehensive}.

Here, we report on a method that significantly improves the printing fidelity of volumetric additive manufacturing in scattering materials. We demonstrate it for tomographic volumetric printing, although the method is also broadly applicable to other light-based volumetric approaches. The idea is to first experimentally characterize the propagation of light through the photo-sensitive material and then to use the most relevant parameters to correct the projected patterns. We propose here to compensate for the attenuation of ballistic light as well as, most importantly, the increased blur of the pattern with depth. We demonstrate the performances of our technique through a series of different centimeter-scale printed objects, in resins whose scattering mean free path is around 5 mm.

\section{Methods}

Scattering of light by random media occurs in many areas of science and technology including imaging, remote sensing or optical communication. The wave disturbance caused by the microscopic refractive index inhomogeneities in such media creates serious hindrance for transmitting or collecting information \cite{ishimaru1978wave}. Many natural environments are scattering such as fog, clouds, a glass of milk or biological tissues. Typically in such disordered media light gets deviated from its straight path when striking an obstacle \cite{bohren2008absorption}. When light is coherent, scattering produces speckle phenomena \cite{goodman2007speckle}. Owing to the availability of spatial light modulators, these interference patterns often seen as a nuisance can now be manipulated thanks to wavefront shaping techniques like feedback-loop optimization scheme \cite{vellekoop2007focusing, horstmeyer2015guidestar} or a transmission matrix approach \cite{popoff2010measuring, boniface2017transmission}. When light is of low-coherence, the speckle loses its contrast and the effect of several scattering events mainly results in the spatial divergence of the beam. To describe this phenomenon, it is often convenient to separate the intensity field propagating through the medium into two components: the ballistic light and the diffuse light. On the one hand, ballistic light experiences no scattering and thus travels straight through the material. Its intensity is attenuated exponentially with penetration depth by both scattering and absorption (Beer-Lambert law). Hereafter, light attenuation by the resin is mainly due to scattering. Its characteristic distance is merely the scattering mean free path, denoted $l_s$, and corresponds to the average distance between two successive scattering events. Another important physical quantity to describe the scattering of light is the transport mean free path $l_t = l_s/(1-g)$, where $g$ stands for the anisotropy coefficient, and represents the length over which the direction of light is randomized. On the other hand, the diffuse intensity is created entirely within the medium through the phenomenon of scattering.  
In the case of volumetric printing, light scattering is very detrimental. It prevents from depositing the desired light dose in precise 3D locations of the resin container. It translates into severe deviations between the actually printed object and the target geometry that is aimed to be achieved. The most visible differences are mainly the presence (respectively absence) of unwanted (respectively wanted) parts and a global loss of resolution. There is a need for taking into account the scattering properties of resin when computing the projected light patterns. In this purpose, it is critical to characterize the complex effects of scattering for improving the printability.

\begin{figure}[t!]
\centering
\includegraphics[width=\linewidth] {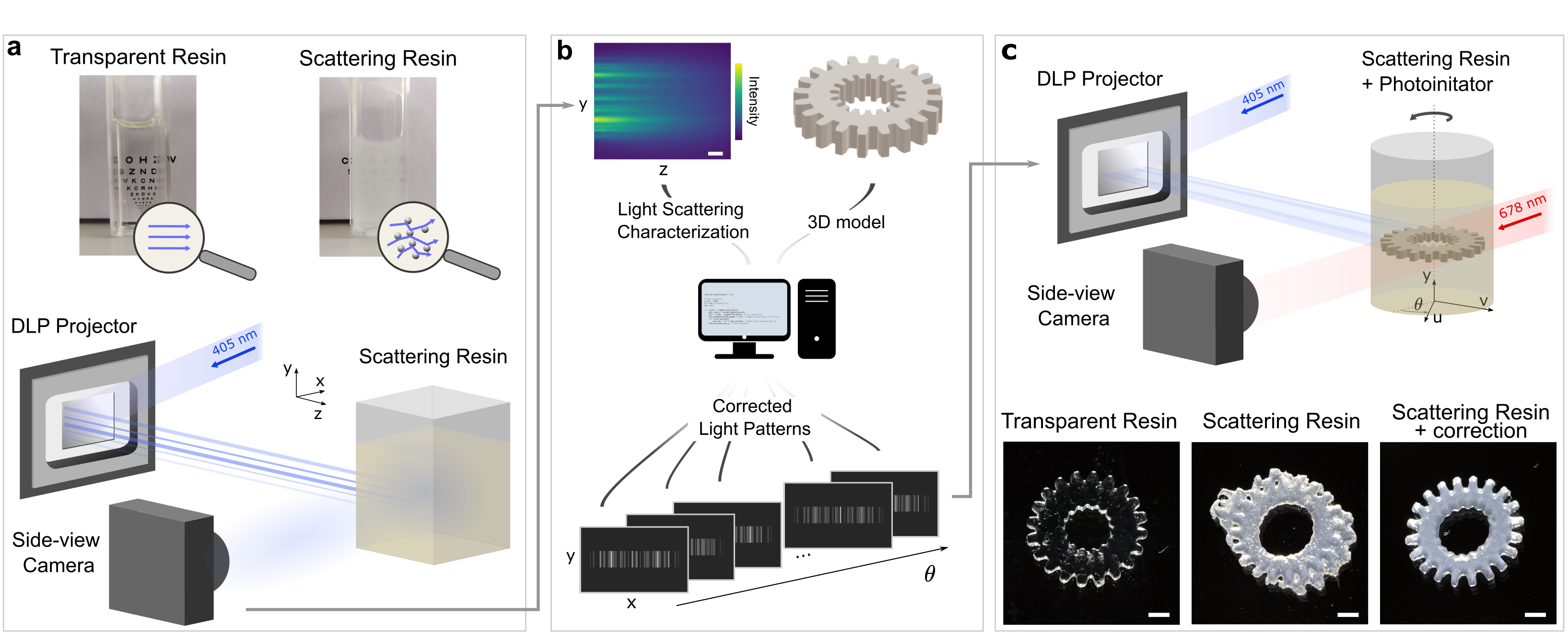}
\caption{ \textbf{Overview of the volumetric printing process in scattering resin}. \textbf{a.} Top. Pictures of transparent and TiO$_2$ loaded acrylate resin in a 10 mm thick square cuvette. Bottom. Experimental setup to characterize light scattering by the resin. A detailed scheme of the setup is provided in Supplementary A. \textbf{b.} Simplified schematic of the framework use to generate the corrected patterns. From the experimentally-acquired images, light scattering is characterized and parameters of interest are used, together with the 3D model of the target print, to compute the scattering corrected patterns (scalebar: 1 mm). \textbf{c.} Representation of a standard tomographic volumetric printer and pictures of the obtained 3D printed parts. (Scalebars: 2 mm).
}
\label{Fig1}
\end{figure}

As sketched in Figure \ref{Fig1}.a, this optical characterization is performed by projecting one or a set of light patterns with a spatial light modulator (here, a Digital Micromirror Device, DMD), through a square cuvette containing the scattering material. The latter is prepared from a transparent resin (here, but not limited to, acrylate resin) in which a controlled amount of TiO$_2$ nanoparticles is homogeneously dispersed (see Materials and Methods section). This protocol allows us to easily adjust the scattering effects on light; we ensure that it is deleterious for volumetric printing. Quantitatively, the scattering resins used herein have a typical scattering mean free path of $l_s \simeq 5$ mm and transport mean free path of $l_t \simeq 8.2$ mm (see Supplementary B). As one can see on the pictures in Figure \ref{Fig1}.a, the impact of scattering is visible to the naked eye. 

The light patterns chosen for the scattering characterization are deliberately narrow along the x-axis (see Figure \ref{Fig1}.a) in order to improve the optical sectioning and increase the contrast of the image obtained on the camera. The position of the square cuvette is aligned such that the projected patterns fall close to its edge. Note that for the moment there is no photo-initiator in the resin.
A camera, perpendicular to the optical axis (i.e. a side-view camera), is used to image the lateral facet of the cuvette (side-view configuration). 
The latter captures only a small part of the scattered light. It is important to emphasize here that the amount of light reaching the camera sensor is intrinsically related to the angular distribution of light scattered by the particles, also referred to as the phase function. In the case of TiO$_2$, the particles are very scattering and light is deviated almost isotropically (see Supplementary B). In contrast, other scattering materials like cells in biological tissues, would exhibit a stronger forward scattering and the probability for light to reach the side-view camera would be much lower. In that situation, a fluorescent dye can be added to the resin and the corresponding isotropic fluorescence signal recorded on the camera.
A typical image obtained with the side-view camera is given in Figure \ref{Fig1}.b. The main effects of scattering on the light intensity profile can be put into two physical phenomena that are: \emph{(i)} an exponential decrease of ballistic light with depth and \emph{(ii)} an increased blur of the light pattern along its propagation.
From these snapshot images, we extract the most relevant physical parameters that describe at best how light is scattered along its propagation. We then take them into account to compute the scattering-corrected light patterns to print the desired target object (for instance, a gear with inner and outer teeth). Finally, these corrected patterns are displayed onto the projector for printing, see Figure \ref{Fig1}.c. During the printing, the side-view camera is used to monitor in real-time the polymerization of the resin, see Supplementary C. The effect of the correction is studied by comparison to prints with standard light patterns, as generated in \cite{kelly2019volumetric,loterie2020high}, in the same scattering resin and in a transparent resin (no TiO$_2$ nanoparticles).
\begin{figure}[t!]
\centering
\includegraphics[width=\linewidth] {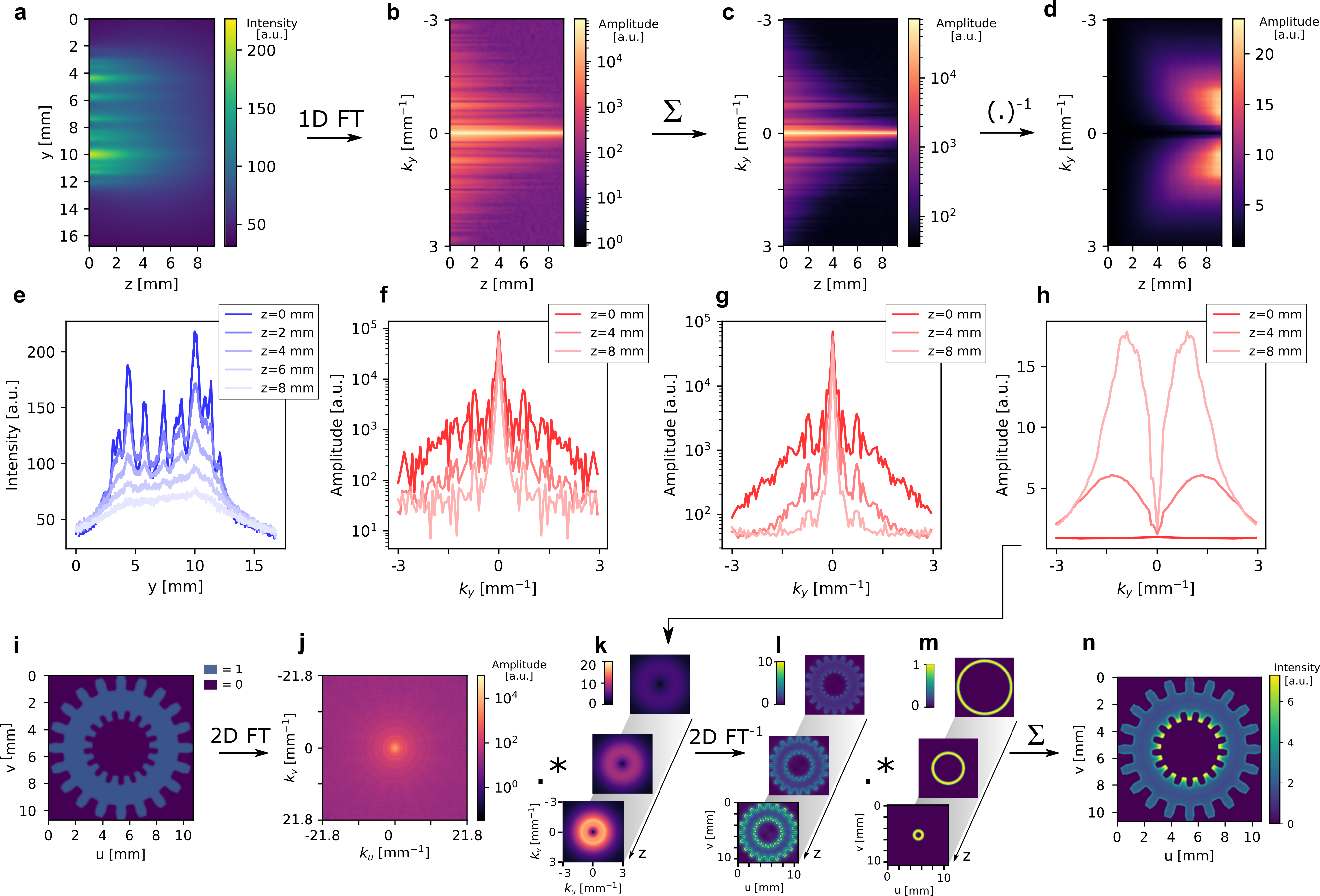}
\caption{ \textbf{Scattering correction workflow}. \textbf{a.} Typical side-view image. \textbf{b.} Corresponding one-dimensional Fourier transform of the pattern along y-axis as a function of penetration depth. \textbf{c.} Sum over a set of different projected patterns (here 100 patterns) \textbf{d.} Correction needed for maintaining a constant energy for all the spatial frequencies at all depths. \textbf{e.-h.} Cross sections of the above images for different depths. \textbf{i.} Binary target 2D object to print. \textbf{j.} Amplitude of the corresponding 2D Fourier transform \textbf{k.} Two-dimensional masks obtained from the interpolation (in polar coordinates) of the curves in (h.) \textbf{l.} Inversed two-dimensional Fourier transform of the filtered space. \textbf{m.} Each filtering is valid only for a given depth, corresponding to a specific annulus in polar coordinates. \textbf{n.} Resulting scattering corrected light dose.
}
\label{Fig2}
\end{figure}

\vspace{1cm}

A challenging task in volumetric tomographic printing is to determine the required light patterns from the desired dose distributions. An interesting aspect of this two-dimensional inverse problem is its close relationship to computed tomography (CT), widely used in medical imaging \cite{kak2002principles}, that aims at reconstructing a three-dimensional image from its projections. Under some simplifying assumptions, these two problems – CT imaging and 3D printing – are in fact mathematically identical. It results that the same algorithm as developed for CT can be successfully applied to volumetric printing. In the present case, the light patterns are calculated using a filtered back-projection algorithm. The conventional approach as described in \cite{kelly2019volumetric,loterie2020high} consists of, first, converting the target 3D model into a three-dimensional binary matrix of voxels (hereinafter designated as binary map), where the entries “1” indicate the presence of matter and “0” its absence at each particular location in space. Then, for each 2D section of this matrix, projections are calculated over multiple angles from the Radon transform \cite{kak2002principlesCompTomographic}. Additional processing are performed to ensure a correct sampling of the projection space and the absence of negative values, that cannot be generated with light (see Supplementary D for further information). 
This approach supposes that light patterns propagate in a straight line without being attenuated or distorted which is no longer valid when printing in non-transparent materials. 

In Figure \ref{Fig2}, we describe our proposed method that takes the effect of scattering into account prior to computing the patterns. 
The fact that light patterns get increasingly blurred with depth (see Figure \ref{Fig2}.a and different z-profiles in Figure \ref{Fig2}.e) is also noticeable in the frequency space (also referred to as “k-space” in the following): the scattering acts as a low-pass filter. In other words, features of high spatial frequencies in the pattern get more rapidly attenuated than the low ones. This is exemplified in Figure \ref{Fig2}.b by computing the one-dimensional Fourier transform of Figure \ref{Fig2}.a along the y-axis (corresponding profiles at different depths are shown in Figure \ref{Fig2}.f). In order to properly characterize the transmission of all the spatial frequencies, a sequence of different patterns (here 100 patterns) is projected onto the DMD. Note that it is important to project patterns whose k-spaces are representative of the frequencies at stake when printing. Suitable sets of patterns include patterns from the Radon transform of the object, random patterns, patterns from the Fourier transform of images or signals, or designed dictionaries of patterns with different spatial frequencies, for example. In Figure \ref{Fig2}.c, the average transmission of the spatial frequencies shows the strong attenuation of the high spatial frequencies as light penetrates in the material. To alleviate this unequal impact of scattering in k-space, the amplitude of the frequency components (which are dampened by the scattering) is enhanced. According to these measurements, one can compute a correction mask that ensures to get the amplitudes of all the incident frequency components (at z = 0 mm) constant across the full vial (at all depth z), see Figure \ref{Fig2}.d. In practice, this correction mask is obtained by dividing the incident averaged spectrum (k-space domain) at z = 0 mm by each spectrum taken at different z depths. Profiles at different z are plotted in Figure \ref{Fig2}.h. They present two symmetric lobes whose amplitude varies with depth. As expected, the correction to apply is more important for high spatial frequencies up to a certain point above which the correction drops, simply because the initial energy (at z = 0 mm) of the corresponding frequencies is very low. These correction masks are 1D whereas the target objects are 2D (slice of the 3D object). Importantly the effect of scattering in k-space should be the same whatever the direction (k$_x$, k$_y$ or k$_z$), simply because the TiO$_2$ particles are homogeneously dispersed in the resin. So there can be easily computed a higher dimensional mask from the curves in Figure \ref{Fig2}.h by applying an axial symmetry with respect to the central frequency k$_y$ = 0 mm$^{-1}$, see Figure \ref{Fig2}.k. The correction mask is then applied onto the target binary object in Fourier (see Figure \ref{Fig2}.j). Different filtering masks are computed depending on the depth at which the correction is performed. The longer the distance over which light must travel in the scattering resin the stronger is the correction to be applied (see Figure \ref{Fig2}.l). 

In our tomographic system, the penetration depth increases radially and is maximal at the center of the vial (i.e. 8 mm in our case), because of the rotation. This region is, in such apparatus, the one where the scattering of light causes the most difficulties and therefore where the correction must be the most important. Because the correction in k-space is depth-dependent, they are valid for restricted regions that look like annulus (see Figure \ref{Fig2}.m). Their radius are connected to the depth at which the correction is done and their thickness gives the accuracy of the reconstruction. For sake of clarity, the depth discretization is of 500 micrometers but a step of 100 micrometers is preferably taken in practice. In Figure \ref{Fig2}.n we represent the resulting scattering corrected light dose, which is the sum over all the annular dose distributions. Compared to the binary map, conventionally use for computing the patterns, this corrected target obtained from experimental measurements has a much higher contrast, especially when moving towards the center.
The latter is then used as target light dose (instead of the binary map Figure \ref{Fig2}.i), to compute the scattering corrected light patterns. The computation of the patterns (with the Radon transform) is performed very similarly to what is currently done with standard tomographic printer (see Supplementary D).

\section{Results}
The performances of the method are investigated experimentally through the printing of different objects. For each, we compare the obtained 3D printed shapes with and without the scattering correction. Importantly, except for the correction applied on the target dose, the whole procedure to retrieve the light patterns based on the Radon transform is exactly the same in both cases (see Supplementary D). In Figure \ref{Fig3}, we assess the gain in print fidelity. 
We use the 3D model of a gear with inner and outer teeth of different size as a resolution target. This object is challenging to print in a scattering resin due to its small features (inner teeth: width = 460 \textmu m, outer teeth: width = 750 \textmu m) which are far from the vial's edge, meaning that light is mostly scattered ($l_s = 6.1$ mm, $l_t = 10$ mm). If no correction is applied, the only way to deposit more light close to the middle of the vial and thus print the inner teeth is to increase the dose. This can be done either by enhancing the laser power or by printing over a longer time. Whereas such increase would bring more light at depth, it will also rise the dose close to the edge of the vial. It results that when the inner teeth start to form, the outer ones are already over-polymerized (middle panel). It is at this level that the correction intervenes to limit the effect of scattering on the print. Instead of computing the light patterns from the binary map (Figure \ref{Fig2}.i), we use the target dose reconstructed from the experimental characterization of light scattering (Figure \ref{Fig2}.n). Corresponding printed gear reported in Figure \ref{Fig3} (bottom panel), show the achieved improvements: the inner teeth are better defined and no over-polymerization of the outer structure is observed. From this 2D+1 object (i.e. with one symmetry axis), we can compute the Intersection over Union of its projection, which is a quantitative way of reporting the printing fidelity\cite{schwab2020printabilityAndShapeFidelityReview} . We report improvements in print fidelity from $IoU = 0.56 \pm 0.02$   to $IoU = 0.80 \pm 0.03$ by printing with a set of corrected tomographic patterns. The baseline print fidelity for this shape in a transparent resin was $IoU = 0.83$. More importantly, applying corrections for scattering allows to fabricate a functional part with protrusions and indentations.

\begin{figure}[t!]
\centering
\includegraphics[width=\linewidth] {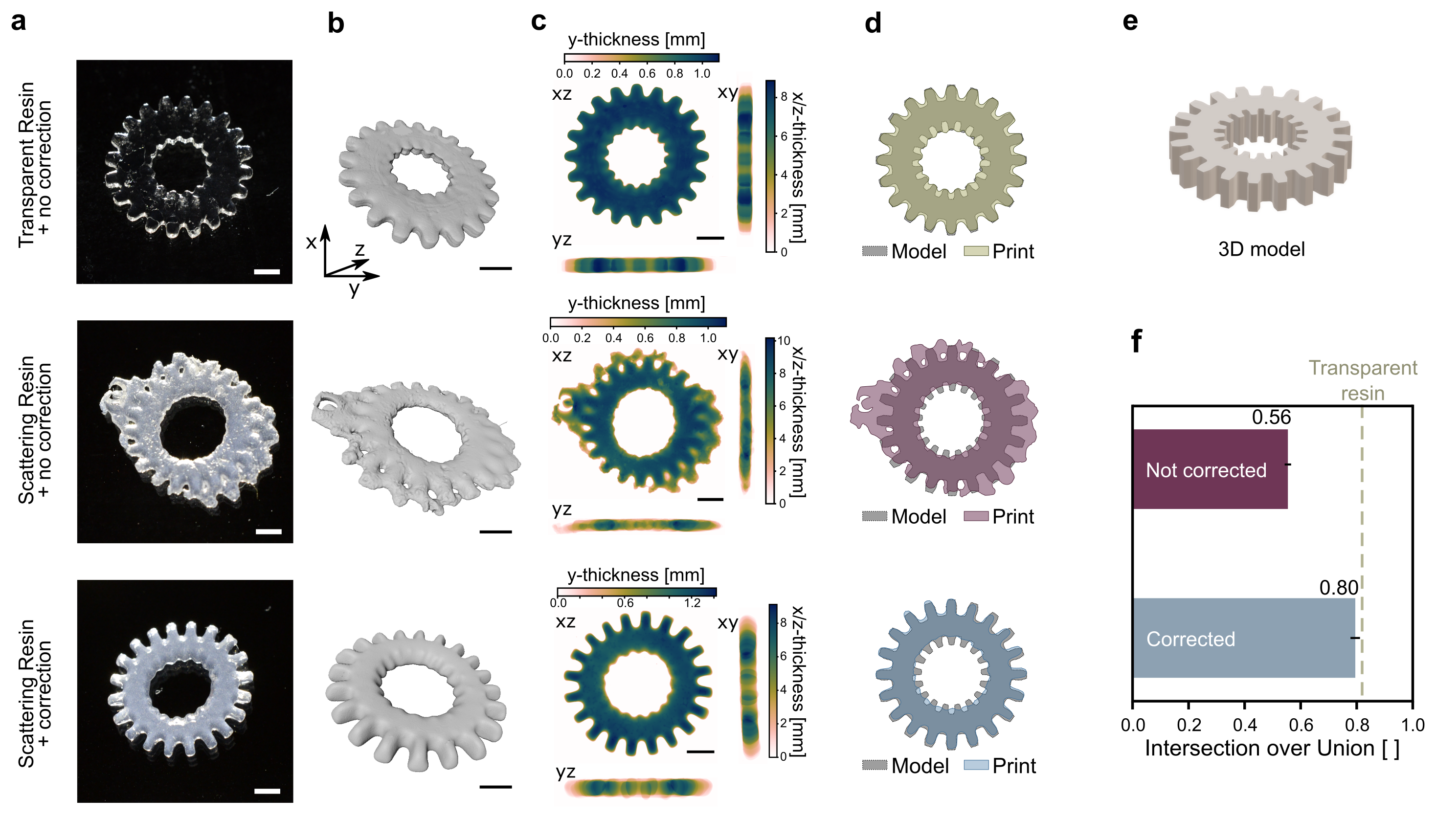}
\caption{\textbf{Quantitative assessment of print quality in scattering resins.} \textbf{a.} Photographs of  printed structures. \textbf{b.} Corresponding x-ray micro CT scan of the printed parts. \textbf{c.} Orthogonal views of thickness profile for the printed part. \textbf{d.} Overlap of the parts and the model. \textbf{e.} 3D model. \textbf{f.}  Intersection over Union between the prints and the model compared to the baseline of printing in a transparent resin; error bars = standard deviation. For scattering resins $l_s = 6.1$ mm, printable depth = 15 mm. Scale bars 2 mm.
}
\label{Fig3}
\end{figure}

\begin{figure}[t!]
\centering
\includegraphics[width=\linewidth] {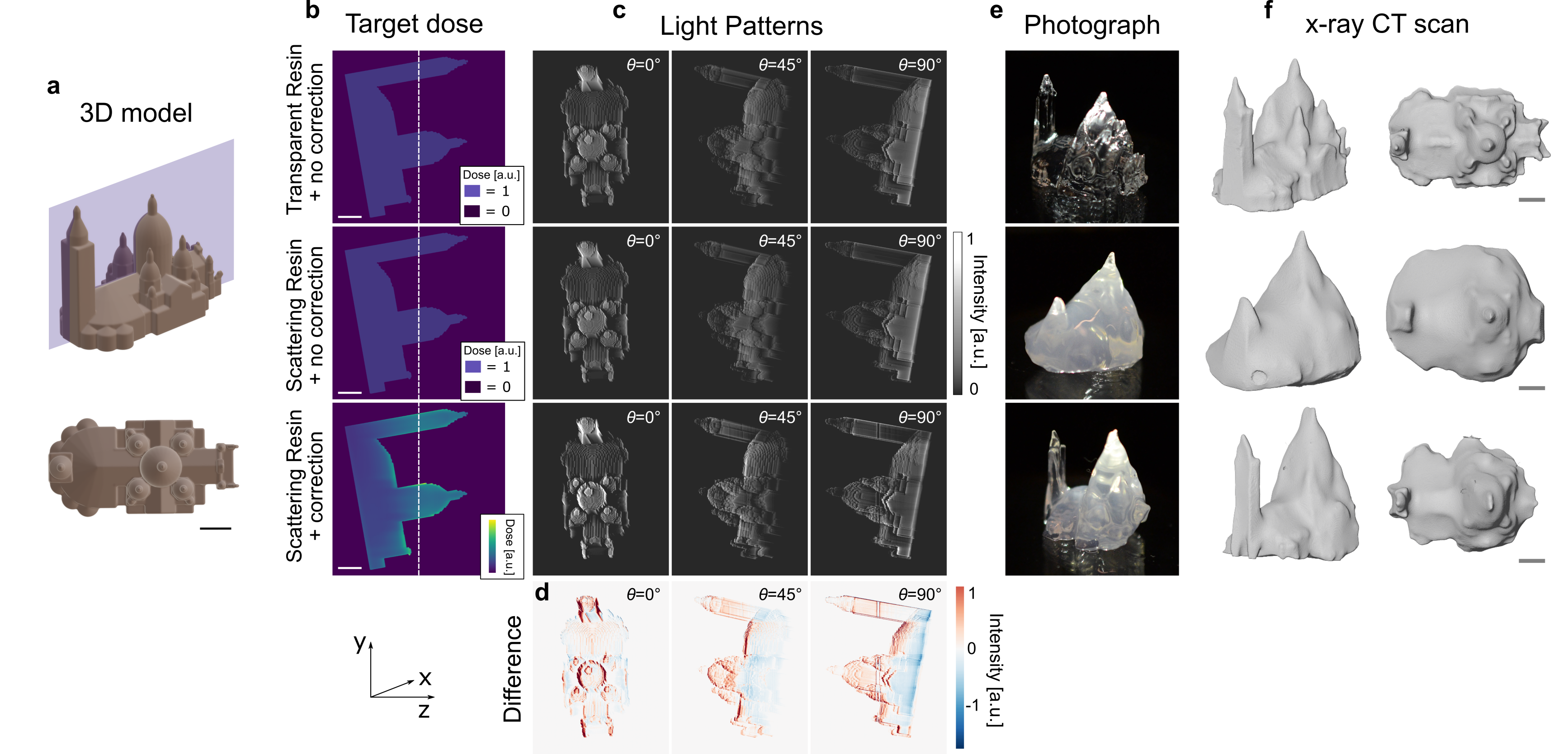}
\caption{\textbf{Volumetric digital manufacturing in scattering resins}. \textbf{a.} 3D model of the Sacré Coeur basilica in Paris. \textbf{b.} Target dose to polymerize the resin. Top and Middle panel: binary mapping (no correction), Bottom panel: Scattering-corrected target dose. \textbf{c.} Corresponding exemplary projected patterns to polymerize the resin for different angles. \textbf{d.} Difference between bottom and middle line to vizualise the resulting correction. \textbf{e.} Photographs of the printed shapes. In this experiment the scattering mean free path is ($l_s = 4.8$ mm, $l_t \simeq 7.9$ mm) \textbf{f.} Oblique and top views from a x-ray micro Computer Tomography (CT) scan of the resulting objects. (Scalebars: 2 mm).
 }
\label{Fig4}
\end{figure}

In Figure \ref{Fig4}, we report on the print of a true volumetric object (with no axial symmetry): the Sacré Coeur basilica in Paris. As in Figure \ref{Fig3}, we compare the prints obtained in a transparent resin with those obtained in a scattering resin ($l_s = 4.8$ mm, $l_t \simeq 7.9$ mm) for the cases where there is no correction and when we apply a correction due to scattering. The correction applied onto the target 3D dose is done following the same protocol as in Figure \ref{Fig2}. A cross-section along the x-axis (indicated in Figure \ref{Fig4}.a) provided in Figure \ref{Fig4}.b shows the importance of depositing more light on the edge of the object (regions of high spatial frequency) which increases the contrast of the target dose and thus compensate for the scattering effects that tends to smooth the print. One can also notice the correction ensures to bring more light close to the middle of the vial (indicated by vertical dashed white lines) to take into account the attenuation. This correction onto the target dose translates in the set of projected patterns to print. In Figure \ref{Fig4}.c we represent the light patterns at $\theta = 0^{\circ}, 45^{\circ}$ and $90^{\circ}$ and shows the induced correction by computing the difference map (see Figure \ref{Fig4}.d). Figure \ref{Fig4}.e and f, are photographs and x-ray scans of the obtained print, respectively. As already observed with the gear in Figure \ref{Fig3}, light scattering prevents from printing the full object without over-polymerizing parts of the structures, mainly the ones closer to the edge of the vial. For instance here, the main square tower cannot be printed correctly. This issue does not occur when the scattering is taken into account prior to computing the light patterns. Also, the correction offers a significantly better resolution to the resulting print. 
Naturally, even with scattering correction, printing in scattering resin has some limitation and the resolution of analogous structures in transparent resins cannot be achieved. Here, the correction aims at depositing more or less light, knowing in advance how light intensity is scattered on average. However, a finer characterization of the scattering process using a laser, including the distortion of its phase for instance, might provide a better correction and would potentially allow, using wavefront shaping techniques, for improving the printing fidelity.

\section{Discussion and Conclusion}
We reported on the necessity of extracting relevant physical quantities that describe the complex effect of scattering and apply in return an appropriate correction prior to printing. We proposed to apply a correction in k-space but other strategies may also be envisioned. We also show how to correct, directly in real space, for the exponential decrease of ballistic light (see Supplementary E). This type of correction is more general since it can also be applied to other materials, like transparent absorptive resins. 
We must emphasize here that the proposed method still relies on the use of ballistic light like the conventional technique (i.e. without scattering corrections). This sets an upper limit in terms of scattering: printing in opaque material where light undergoes multiple scattering is not feasible. However, the scattering regime studied here, where the correction proves to improve the printing fidelity, is still very relevant for many interesting materials. For instance, cell-laden biological resins  currently used to synthesize living tissues are scattering \cite{moroni2018biofabrication}. Compared to TiO$_2$ nanoparticles, cells and other biological compounds are much less scattering \cite{jacques2013optical}, and printing in them with no scattering correction is possible but limited to either low concentration of cells (around 10$^7$ cells/mL) or small shapes (centimeter-scale) \cite{bernal2019volumetric}. With the ability of taking the scattering of light into account prior to computing the light patterns, we strongly believe that cell concentration in hydrogels can be increased without impacting the prints. 

In summary, we presented a simple yet effective strategy to significantly improve the performance of tomographic volumetric printers with scattering resins. The idea relies on characterizing the effect of scattering on the propagation of light and correcting accordingly the projected patterns.  An advantage is that the experimental characterization can be performed on the printer directly using a side-view camera, orthogonal to the optical axis. Corresponding captured images reveal the poor transmission of high spatial frequencies caused by the  scattering events. The numerical correction aims at compensating for this frequency-dependent attenuation by accentuating the features of highest spatial frequencies. The resulting target light dose has an increased contrast compared to the standard binary map, traditionally used. Through the printing of different 3D objects, this correction proves to be essential for improving the fidelity between the 3D model and the printed shape. 
Although demonstrated on a tomographic volumetric printer, similar corrections could also be applied to other printing technologies, such as stereolithography, DLP printing, and also to more recent techniques like xolography \cite{regehly2020xolography}, two photon fabrication, longitudinal or multi-axial setups.

\section*{Acknowledgments}
The authors would like to acknowledge Gary Perrenoud and Edward Andò for their support with the micro CT imaging of the printed structures.

\bibliographystyle{IEEEtran}
\bibliography{scifile}

\newpage

\section*{Materials and Methods}

\subsubsection*{Volumetric tomographic 3D printer}
\begin{figure}[htbp]
    \centering
    \includegraphics[width=\linewidth]{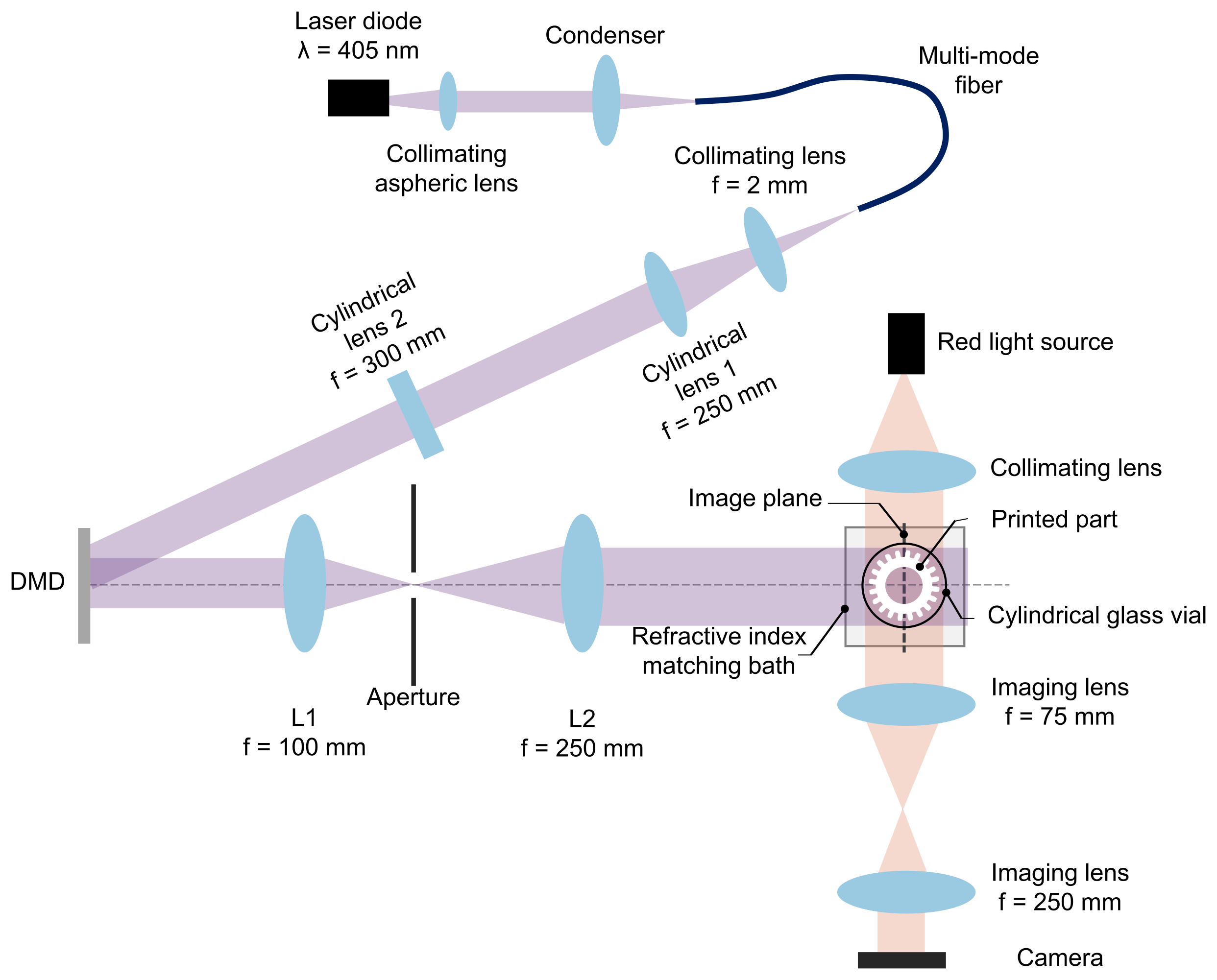}
    \caption{Optical setup of the tomographic volumetric 3D printer.}
    \label{fig_S_OpticalSetup}
\end{figure}

The optical setup for tomographic additive manufacturing is depicted in figure \ref{fig_S_OpticalSetup}. Blue light from 4 continuous laser diodes at 405 nm (HL40033G, Ushio, Japan) was condensed into a multimode optical fiber (WF 70×70/115/200/400N, CeramOptec, Germany) by means of aspheric lenses (C671-TME405, Thorlabs, USA). Light was then collimated by means of two cylindrical telescopes onto a Digital Micromirror Device (VIS-7001, Vialux, Germany). Light patterns from the DMD are then projected onto the resin by means of lens pair with focal lengths $f_{1} = 100$ mm (AC254-100-A-ML, Thorlabs) and $f_{2} = 250$ mm (ACT508-250-A-ML, Thorlabs). An iris at the common focal plane of the lenses filters out high diffracting orders from the DMD.

The photosensitive resins are held in cylindrical glass vials (diameter 16.5 mm). These vials are set to turn with a high-precision rotary stage (X-RSW60C, Zaber, Canada). 

Orthogonally to the optical axis of the printer, red light at 678 nm from a laser diode is used to image the printing process. A lens pair with focal lengths $f_{1} = 75$ mm (AC508-075-A-ML, Thorlabs) and $f_{2} = 250$ mm (ACT508-250-A-ML, Thorlabs) produces an image onto a CMOS camera (ACE ACA2000-50G, Basler, Germany).

\subsubsection*{Acrylic resins}
The photo-curable resin used in this work was prepared by combining di-pentaerythritol pentaacrylate (SR399; Sartomer, France) or PRO21905 (Sartomer, France) with 0.6 mM phenylbis(2,4,6-trimethylbenzoyl)phosphine oxide (97  \%; Sigma Aldrich, USA) in a planetary mixer (KK-250SE, Kurabo, Japan). A threshold light dose is necessary to induce solidification of the liquid resin. This threshold depends on the functionality of the resin \cite{loterie2020high} and on oxygen inhibition \cite{ligon2014OxygenInhibition}.
To make the resins scattering, TiO2 nanoparticles ($<100$ nm particle size, 99.5 \%, Sigma Aldrich, Switzerland) were first diluted in ethanol ( 99.8 \%, Fischer Chemical, South Africa ) and then added to the resins before planetary mixing. The calculated scattering phase functions for these resins is depicted in Supplementary Figure \ref{FigureSuppScattProp}. The phase function was calculated using the online tool \url{https://omlc.org/calc/mie_calc.html}.
The resins were poured into cylindrical glass vials and sonicated for 15 minutes to remove bubbles.

\subsubsection*{Characterization of the scattering profile of resins}
A small amount of the resins was put aside before adding the photoinitiator and poured into 10mm cuvettes with 4 polished windows. The cuvettes were placed at the image plane of the printer. 
Series of patterns were displayed on the DMD while photographs were recorded simultaneously with the orthogonal camera. 

\subsubsection*{Post-processing of printed parts}
Parts were post-processed by rinsing them in isopropyl alcohol (99 \%, Thommen-Furler, Switzerland) for 3 minutes under sonication.  

\subsubsection*{MicroCT imaging and assessment of print fidelity}

Printed objects were imaged under a 160 kV X-ray transmission tomograph (Hamamatsu, Japan) with voxel sizes of 8.4 \textmu m x 8.4 \textmu m x 8.4 \textmu m. 3D visualizations of the pieces were obtained with Avizo software (ThermoFischer, USA). 

Quantitative analysis of 3D scans were performed on ImageJ \cite{rueden2017imagej2}. To quantify print fidelity, the object in the microCT scan data was segmented and binarized using Otsu's thresholding \cite{otsu1979threshold}. The images of the object were centered around its center of mass and rotated to align them with the orientation of the reference shape. The processed stack of images was then saved and imported into a python code, which automatically computed the Intersection over Union (IoU) for several affine transformations (excluding shear) of the image. From this, we obtained the distribution of IoU indices for each part. We reported the mean IoU and its standard deviation.

\subsubsection*{3D models}
We used FreeCAD (\url{https://www.freecadweb.org/}) to design the 3D models for the gears and the twisted rings. 3D models for the Sacré Coeur Basilica were obtained freely from \url{https://www.thingiverse.com/thing:311002/files} and designed by \textit{LeFabLab}.

\newpage

\section*{Supplementary}

\subsubsection*{A. Tomographic volumetric 3D printer}

\subsubsection*{B. Optical properties of the scattering resins}
\begin{figure}
    \centering
    \includegraphics[width=\linewidth]{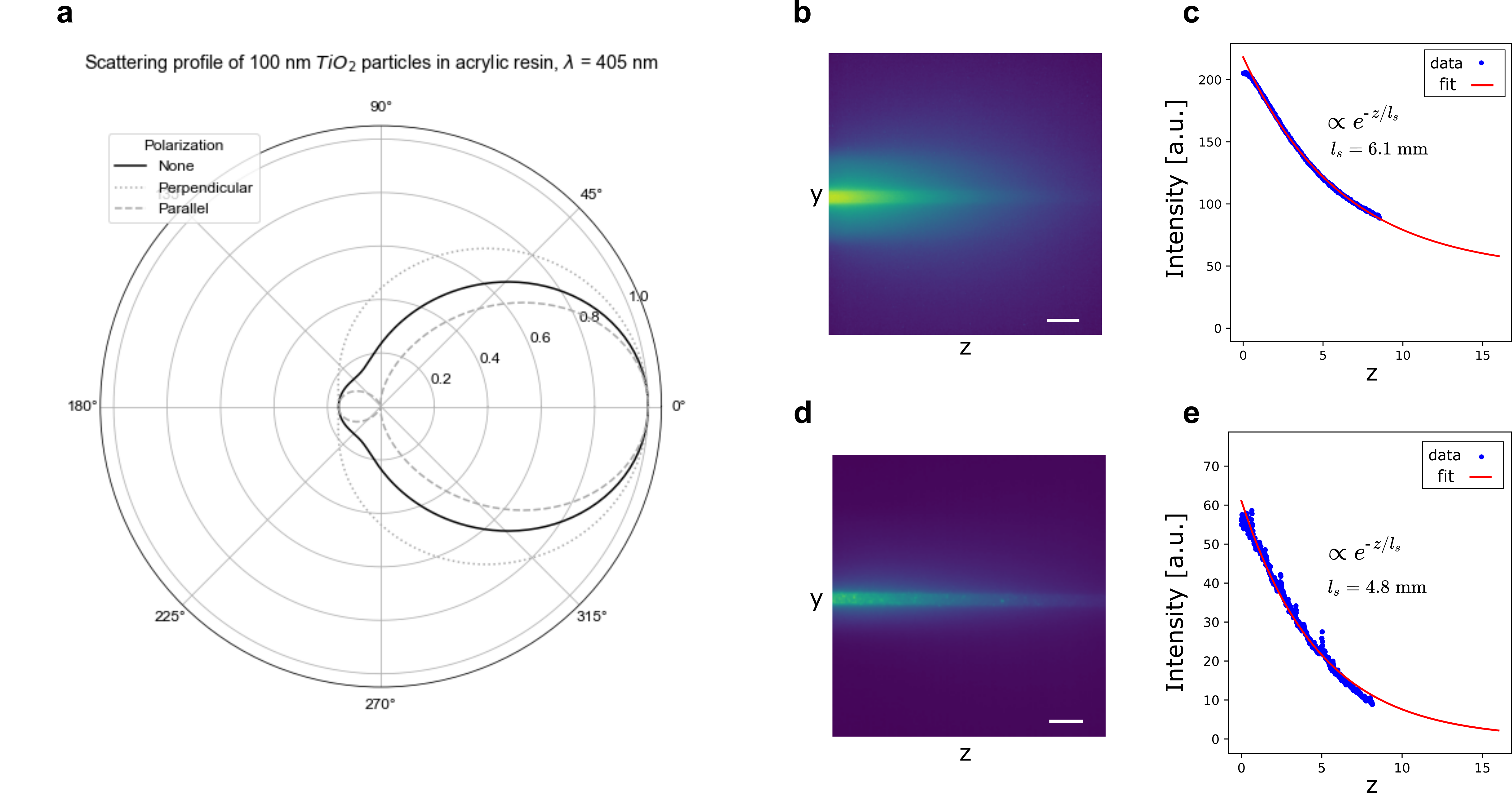}
    \caption{\textbf{Optical properties of the scattering resins.} \textbf{a.} Phase function of TiO$_2$ at $\lambda = 405$ nm. \textbf{b.} Side-view image of a thin sheet of light through the resin used in Figure \ref{Fig3}. \textbf{c.} Exponential decrease of ballistic light. \textbf{d.} Side-view image of a thin sheet of light through the resin used in Figure \ref{Fig4}. \textbf{e.} Exponential decrease of ballistic light.}
    \label{FigureSuppScattProp}
\end{figure}

In this section we provide additional information on the scattering properties of our resins. The scattering resins are prepared in the following manner: a transparent resin (here, Dipentaerythritol Pentaacrylate, SR399, Sartomer Arkema) is made scattering by adding TiO$_2$ nanoparticles (TiO$_2$ nanopowder, $<100$ nm particle size, Sigma Aldrich) homogeneously dispersed in the resin. The concentration of TiO$_2$ is around 0.3 mg/mL. Although this concentration is low, the scattering induced by TiO$_2$ nanoparticles is very high (refractive index of TiO$_2$ is 2.9 at 400 nm compared to 1.5 for the monomer). Knowing the size of the particles and the refractive index mismatch we can derive from Mie scattering formulas the theoretical phase function (open source calculator: \url{https://omlc.org/calc/mie_calc.html}). This reveals the relatively high isotropy of scattered light, see Figure \ref{FigureSuppScattProp}.a. It translates into a low anisotropy factor $g = 0.39$ (average cosine of phase function). Experimentally, from the ballistic light exponential decay we can retrieve the scattering mean free path of light, $l_s$. Figure \ref{FigureSuppScattProp}.b-c (resp. Figure \ref{FigureSuppScattProp}.d-e) corespond to the scattering resin used to print the gear in Figure \ref{Fig3} (the Sacré Coeur basilica in Figure \ref{Fig4}). From the scattering mean free path, one can estimate the transport mean free path $l_t = l_s/(1-g)$

\subsubsection*{C. Correction impact on the printing process}
\begin{figure}[htbp]
\centering
\includegraphics[width=\linewidth] {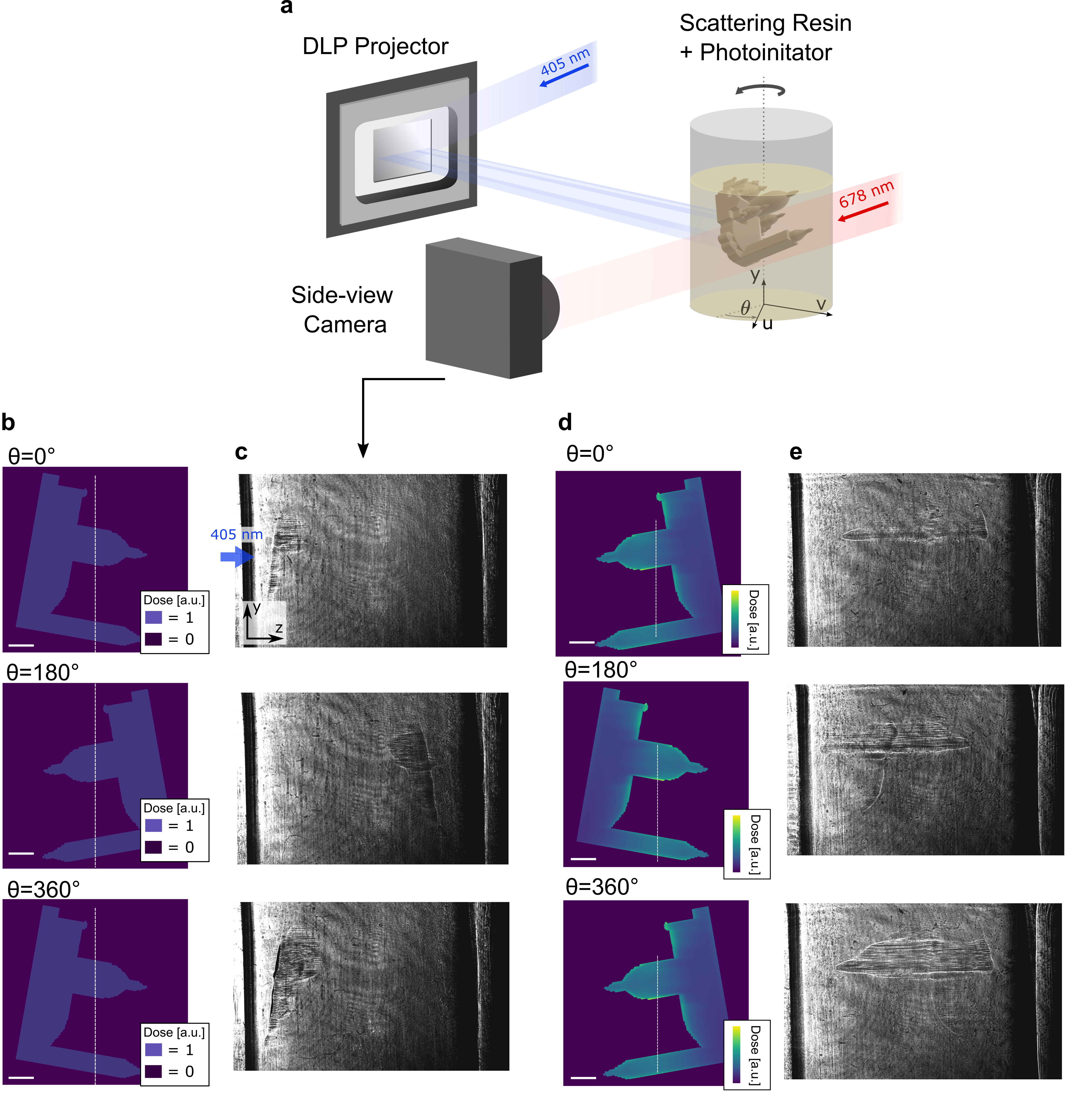}
\caption{\textbf{Real-time monitoring of the printing with and without corrected patterns.} \textbf{a.} Schematic view of the printer with the passive monitoring arm. \textbf{b.} Non-corrected target light dose used to compute the pattern. \textbf{c.} Images acquired on the side-view camera during printing in a scattering resin. When no correction is applied the polymerization starts close to the vial's edge. \textbf{d.} Scattering-corrected target light dose used to compute the pattern. \textbf{e.} Images acquired on the side-view camera during printing in a scattering resin. The applied correction also modifies the printing process. 
 }
\label{FigSuppMonitor}
\end{figure}

To demonstrate that the applied correction changes notably the printing process, we image with the side-view camera images the vial under red light exposure at different times of the polymerization, as shown in Figure \ref{FigSuppMonitor}.a. These videos prove that the correction takes well into account the effect of scattering. If no correction is applied (patterns computed from the binary target dose in \ref{FigSuppMonitor}.b), the object builds from the edge of the vial, where it receives more light. This means printing in the center of the vial requires longer exposure which inevitably over-polymerizes the parts of the object close to the vial's edge. Here, the positioning of the print induces over-polymerization of the basilica's floor as one can see in Figure \ref{Fig4}. The correction prevents from printing the structure radially from the edge to the center of the vial. In Figure \ref{FigSuppMonitor}.e, images acquired at the beginning of the polymerization shows that the dome of the basilica as well as its floor appear at the same time. It demonstrates that the correction apply to deposit more light in the central region of the vial works which strongly limits the over-polymerization of parts of the object.

\newpage

\subsubsection*{D. Light patterns computation from the target dose}
Hereinafter, the computation of the 2D light patterns is explained in further detail. As a difference from standard approaches used in \cite{kelly2019volumetric, loterie2020high, bernal2019volumetric}, the input of the algorithm is not the binary model of the 3D object, but rather the light dose (intensity values) one would like to deposit inside the vial. The two approaches would provide similar results in a transparent resin (assuming light propagates in a straight line and is not attenuated along its propagation). This is no longer true when printing in scattering materials. 
A simplified workflow of the algorithm is sketched in Figure \ref{FigSuppAlgo}. From the target light dose (here attenuation-correction, see Supplementary C), the Radon transform is performed. The obtained sinogram is filtered with respect to its Fourier transform to compensate for the oversampling of the low spatial frequencies. If this filtering operation is not done, the cumulative dose projected into the resin would be blurred and compromise printing resolution. This step generates negative values that cannot be optically implemented. They are set to zero, which creates some artefacts. An optimization is generally run to improve the fidelity between the target dose and the sinogram backprojection. This procedure is standard in tomographic volumetric printing. We follow it strictly but put more efforts in the final optimization. Indeed, the target to achieved according to the present invention is not binary but real positive, which makes the optimization a bit more complex. Therefore, the forward model is modified and the number of iterations increased, but the gradient descent still provides a significant improvement of the patterns. The code is written in python and uses the PyTorch library, so that it can be run on a GPU.

\begin{figure}[htbp]
\centering
\includegraphics[width=\linewidth] {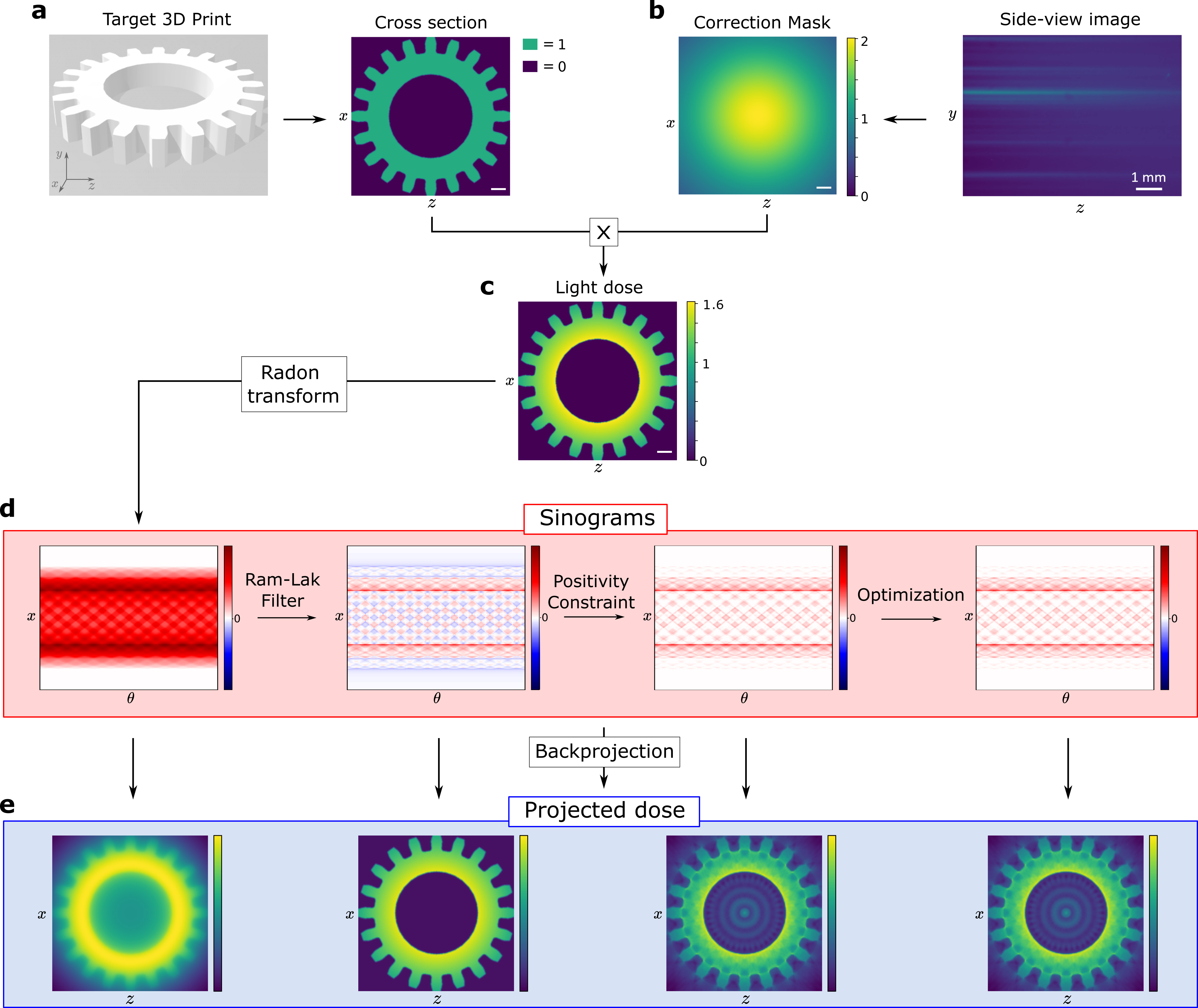}
\caption{\textbf{Procedure to compute the light patterns from the target dose.} \textbf{a.} The target 3D print is cut in different slices along the y-axis. \textbf{b.} From the experimental side-view images we can compute the correction mask (here the attenuation-correction mask). \textbf{c.} Applied onto the binary map, it provides the target light dose that takes into account the effect of scattering (here, the attenuation of ballistic light). \textbf{d.} As in any tomographic volumetric printing technique, we must then compute from the target dose cross section (2D) the set of light patterns (here in 1D). This is achieved by computing the Radon transform. Because the Fourier space is not properly sampled, one needs to apply a Ram-Lak filter \cite{kak2002principlesCompTomographic} . The corresponding patterns contain positive and negative values that cannot be optically generated. All the negative values are set to 0 and an optimization algorithm is designed to optimize around this positivity constraint and maximize the dose fidelity. \textbf{e.} Corresponding reconstructed dose for the different forward models.
 }
\label{FigSuppAlgo}
\end{figure}

\newpage

\subsubsection*{E. Attenuation-correction}
\begin{figure}[htbp]
\centering
\includegraphics[width=\linewidth] {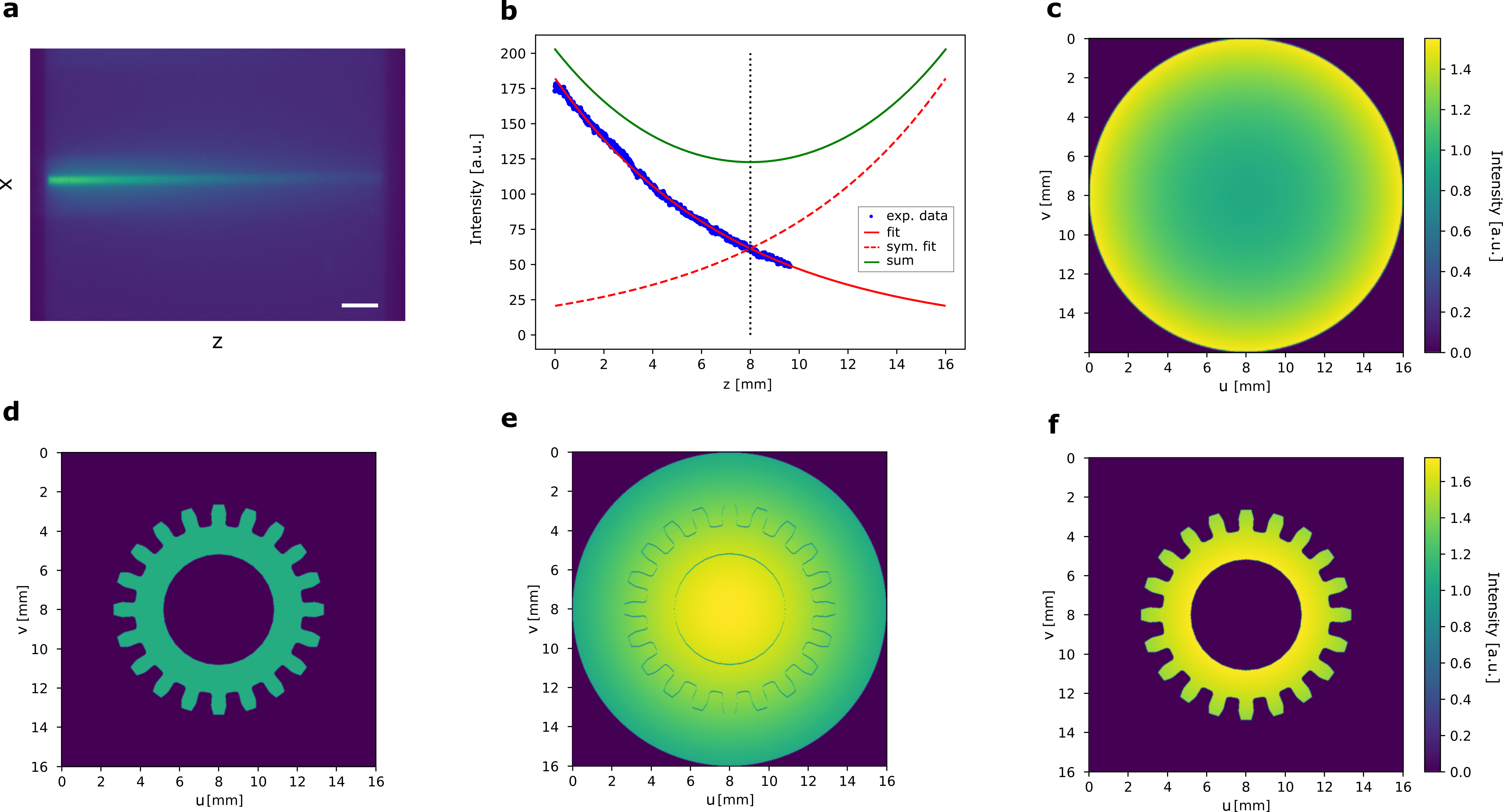}
\caption{\textbf{Attenuation correction of light.} \textbf{a.} Typical side-view image captured on the camera perpendicular to the optical axis. A thin sheet of light is sent through the scattering resin for characterization. Scale bar = 1 mm. \textbf{b.} Experimental data are fitted with a negative exponential. After half a turn the equivalent decrease can be observed from the other side (represented by the symmetrical with respect to the center of the vial at z = 8 mm, dashed red curve). The cumulated dose inside resin after 360$^{\circ}$ is the green curve (i.e. the sum of the two red curves). In average there is a lack of light in the middle of the vial. \textbf{c.} Two-dimensional map of the light amount in the vial (top view). \textbf{d.} Binary mapping of the object to print. \textbf{e.} Two-dimensional correction mask. \textbf{f.} Attenuation-corrected light dose.
 }
\label{FigSuppAtt}
\end{figure}

One main consequence of light scattering is the exponential decrease of ballistic light intensity. This phenomenon can be easily registered and analysed using a detection orthogonal to the optical axis (with the side-view camera presented in Figure \ref{Fig1}). In Figure \ref{FigSuppAtt} we propose a procedure (similar to Figure \ref{Fig2}) to correct for this exponential attenuation of ballistic light. The measurement consists here of sending through the resin a relatively narrow laser beam (in both x and y directions). In practice, this is achieved by activating only a few pixels on the DMD. 
The amount of ballistic light as a function of depth is characterized by measuring the intensity decrease along the z-axis. Experimental data are reported in Figure \ref{FigSuppAtt}.b and fitted with a negative exponential. Fitting coefficients provide an estimation of the scattering mean free path $l_s$, which is the average mean free path length between two successive scattering events. The fit reported in Figure \ref{FigSuppAtt}.b gives $ls = 7.3$ mm. The cuvette used for the measurement is only 10 mm thick but the fit allows extrapolating the trend across the entire vial (cylindrical 16 mm diameter reservoir used for printing). In tomographic volumetric printing, the vial (i.e. the resin container) rotates with respect to its center, and it is thus necessary to consider the total amount of light deposited inside after a full rotation (360$^{\circ}$). In particular after half a turn a similar exponential decrease of ballistic light would be observed, because the TiO$_2$ particles are dispersed uniformly inside the resin. So in average the amount of ballistic light in the vial is the sum of the two negative exponential curves (bold-line curve and its symmetric with respect to the axis of rotation at z = 8 mm in dashed). Because the decrease is exponential, the amount of light in the vial is not uniform after a full rotation. In particular, the proportion of ballistic photons is smallest at the center of rotation of the resin container. Here for the scattering resin under study, there is, in average, 40$\%$ less light in the middle of the vial (i.e. after 8 mm of propagation) than on the edges. In Figure \ref{FigSuppAtt}.c the corresponding two-dimensional profile of ballistic light across the vial (16 mm in diameter) is reconstructed. The latter is obtained by interpolating the bold-line curve reported in the plot of Figure \ref{FigSuppAtt}.b in polar coordinates. This two-dimensional map is then inverted and gives access to what is designated herein as “correction mask” (Figure \ref{FigSuppAtt}.e). To account for the exponential decrease of ballistic light inside the scattering resin, this correction mask is applied onto the binary map (Figure \ref{FigSuppAtt}.d) prior to computing the light patterns with the Radon transform. The resulting light dose to project inside the resin to print, taking into consideration the attenuation of ballistic light, is represented in Figure \ref{FigSuppAtt}.f. It corresponds to the product of the binary map with the correction mask. 

Note that here the correction compensates for the exponential decrease of ballistic light due to scattering resin but the effect of pure sample absorption can be treated similarly. A useful application would be to correct light absorption from dyes or one or more photoinitiators. Absorption from photoinitiators is essential to polymerize the resin, but it limits the performance of the printer, such as for resolution or print size. Usually, the concentration of one or more photoinitiators is chosen so that absorption is very small across the vial, but this means that more light (i.e, more time) is needed to print. Also, correcting for the absorption one or more photoinitiators offers the possibility to print faster, to print in weakly polymerizing or crosslinking materials, or to produce larger objects.


%



\end{document}